\documentclass[twocolumn,showpacs,amsmath,amssymb,showkeys]{revtex4}

\usepackage{graphicx}
\usepackage{dcolumn}
\usepackage{bm}
\usepackage{rotating}
\usepackage{verbatim}
\usepackage{longtable}

\makeatletter 
\gdef\@ptsize{0}
\let\@currsize\normalsize 
\makeatother 
\usepackage{setspace} 
\doublespace 

\usepackage{xcolor}
\usepackage[bookmarks=true,colorlinks,linkbordercolor=blue,linkcolor=blue,citecolor=blue,urlcolor=blue]{hyperref}

\begin{document}

\title{Rock around the Clock:\\An Agent-Based Model of Low- and High-Frequency Trading\footnote{We are grateful to Sylvain Barde, Francesca Chiaromonte, Antoine Godin, Alan Kirman, Nobi Hanaki, Fabrizio Lillo, Frank Westerhoff, for stimulating comments and fruitful discussions. We also thank the participants of the Workshop on Heterogeneity and Networks in (Financial) Markets in Marseille, March 2013, of the EMAEE conference in Sophia Antipolis, May 2013, of the WEHIA conference in Reykiavik, June 2013, of the 2013 CEF conference in Vancouver, July 2013 and of the SFC workshop in Limerick, August 2013 where earlier versions of this paper were presented. All usual disclaimers apply. The authors gratefully acknowledge the financial support of the Institute for New Economic Thinking (INET) grant \#220, ``The Evolutionary Paths Toward the Financial Abyss and the Endogenous Spread of Financial Shocks into the Real Economy''.}}


\author{Sandrine Jacob Leal}\affiliation{CEREFIGE, ICN Business School, Nancy-Metz (France)}\altaffiliation[Also: ]{GREDEG, Sophia-Antipolis (France)}\email[E-mail: ]{sandrine.jacob-leal@icn-groupe.fr} 

\author{Mauro Napoletano}
\affiliation{OFCE, Skema Business School, Sophia-Antipolis (France)}
\altaffiliation[Also: ]{Scuola Superiore Sant'Anna, Pisa (Italy)}\email[E-mail: ]{mauro.napoletano@sciencespo.fr} 

\author{Andrea Roventini}\affiliation{Universit{\`a} di Verona,  Verona (Italy)}
\altaffiliation[Also: ]{Scuola Superiore Sant'Anna, Pisa (Italy) and OFCE, Sophia-Antipolis (France)}\email[E-mail: ]{andrea.roventini@univr.it} 

\author{Giorgio Fagiolo}%
\affiliation{Istituto di Economia, Scuola Superiore Sant'Anna, Pisa (Italy)}\email[E-mail: ]{giorgio.fagiolo@sssup.it} 


\begin{abstract}
\noindent \textbf{Abstract. }We build an agent-based model to study how the interplay between low- and high-frequency trading affects asset price dynamics. Our main goal is to investigate whether high-frequency trading exacerbates market volatility and generates flash crashes. In the model, low-frequency agents adopt trading rules based on chronological time and can switch between fundamentalist and chartist strategies. On the contrary, high-frequency traders activation is event-driven and depends on price fluctuations. High-frequency traders use directional strategies to exploit market information produced by low-frequency traders. Monte-Carlo simulations reveal that the model replicates the main stylized facts of financial markets. Furthermore, we find that the presence of high-frequency trading increases market volatility and plays a fundamental role in the generation of flash crashes. The emergence of flash crashes is explained by two salient characteristics of high-frequency traders, \textit{i.e.,} their ability to i) generate high bid-ask spreads and ii) synchronize on the sell side of the limit order book. Finally, we find that higher rates of order cancellation by high-frequency traders increase the incidence of flash crashes but reduce their duration.

\end{abstract}





\maketitle
\section{Introduction}
\label{sec:intro}

This paper builds an agent-based model to study how high-frequency trading affects asset price volatility as well
as the occurrence and the duration of flash crashes in financial
markets. 

The increased frequency and severity of flash crashes and the high volatility of prices observed in financial time series
have recently been associated to the rising importance of high-frequency
trading \cite{sornette2011crashes}. However, the debate in the literature about the benefits and costs of high frequency trading (HFT henceforth) has not been settled yet. On the one hand, some works stress that high-frequency traders may
play the role of modern market-makers, providing an almost continuous flow of liquidity \cite{menkveld2013high}. Moreover, HFT reduces transaction costs and favors price discovery and market efficiency by strengthening the links between different markets \cite{brogaard2010high}. On the other hand, many empirical and theoretical studies raise concerns about the threatening effects of HFT on the dynamics of financial markets. In particular, HFT may lead to more frequent periods of illiquidity, possibly leading to the emergence of flash crashes
\cite{kirilenko2011flash}. Furthermore, HFT may exacerbate market
volatility \cite{zhang2010high, hanson2011hft} and negatively affect market efficiency \cite{wah2013latency}. 

This work contributes to the current debate on the impact of HFT on asset price dynamics by developing an agent-based model of a limit-order book (LOB) market \footnote{See, for instance, Refs. \cite{farmer2005predictive,slanina2008critical} for detailed studies of the effect of  limit-order book models on market dynamics.} wherein heterogeneous high-frequency (HF) traders interact with low-frequency (LF) ones. Our main goal is to study whether HFT helps to explain the emergence of flash crashes and more generally periods of higher volatility in financial markets. Moreover, we want to shed some light on which distinct features of HFT are relevant in the generation of flash crashes and affect the process of price-recovery after a crash.

In the model, LF traders can switch between fundamentalist and chartist strategies according to their profitability. HF traders adopt \textit{directional} strategies that exploit the price and volume-size information produced by LF traders \cite{securities2010equity,aloud2011directional}. Moreover, in line with empirical evidence on HFT \cite{easley2012volume}, LF trading strategies are based on \textit{chronological} time, whereas those of HF traders are framed in \textit{event} time \footnote{As noted by ref. \cite{easley2012volume}, HFT requires the adoption of algorithmic trading implemented through computers which natively operate on internal event-based clocks. Hence, the study of HFT cannot be reduced to its higher speed only, but it should take into account also the associated new trading paradigm. See also Ref. \cite{aloud2012modelling} for a modeling attempt in the same direction.}. Consequently, LF agents, who trade at exogenous and constant frequency, co-evolve with HF agents, whose participation in the market is endogenously triggered by price fluctuations. Finally, consistent with empirical evidence \cite{kirilenko2011flash}, HF traders face limits in the accumulation of open positions.

So far, the few existing agent-based models dealing with HFT have mainly treated HF as zero-intelligence agents with an exogenously-given trading frequency \cite{bartolozzi2010multi, hanson2011hft}. However, only few attempts have been made to account for the interplay between HF and LF traders \cite{paddrik2011, aloud2012modelling, wah2013latency}. We improve upon this literature along several dimensions. First, we depart from the zero-intelligent framework by considering HF traders who hold event-based trading-activation rules, and place orders according to observed market volumes, constantly exploiting the information provided by LF traders. Second, we explicitly account for the interplay among many HF and LF traders. Finally, we perform a deeper investigation of the characteristics of HFT that generate price downturns, and of the factors explaining the fast price-recovery one typically observes after flash crashes.

We study the model in two different scenarios. In the first scenario (``only-LFT'' case), only LF agents trade with each other. In the second scenario (our baseline), both LF and HF traders co-exist in the market. The comparison of the simulation results generated from these two scenarios allows us to assess the contribution of high-frequency trading to financial market volatility and to the emergence of flash crashes. In addition, we perform extensive Monte-Carlo experiments wherein we vary the rate of HF traders' order cancellation in order to study its impact on asset price dynamics.

Monte-Carlo simulations reveal that the model replicates the main stylized facts of financial markets (\textit{i.e.,} zero autocorrelation of returns, volatility clustering, fat-tailed returns distribution) in both scenarios. However, we observe flash crashes together with high price volatility only when HF agents are present in the market. Moreover, we find that the emergence of flash crashes is explained by two salient characteristics of HFT, namely the ability of HF traders \textit{(i)} to grasp market liquidity leading to high bid-ask spreads in the LOB; \textit{(ii)} to synchronize on the sell-side of the limit order book, triggered by their event-based strategies. Furthermore, we observe that sharp drops in prices coincide with the contemporaneous concentration of LF traders' orders on the buy-side of the book. 
In addition, we find that the fast recoveries observed after price crashes result from both a more equal distribution of HF agents on both sides of the book and a lower persistence of HF agents' orders in the LOB. Finally, we show that HF agents' order cancellations have an ambiguous effect on price fluctuations. On the one hand, high rates of order cancellation imply higher volatility and more frequent flash crashes. On the other hand, they also lead to faster price-recoveries, which reduce the duration of flash crashes. 

Overall, our results validate the hypothesis that HFT exacerbates asset price volatility, generates flash crashes and periods of market illiquidity (as measured by large bid-ask spreads). At the same time, consistent with the recent academic and public debates about HFT, our findings highlight the complex effects of HF traders' order cancellation on price dynamics \footnote{See Ref. \cite{hasbrouck2009technology} for an empirical investigation of the importance of order cancellation in current financial markets and of its determinants. See, among others, the articles in Refs. \cite{EconomistHFT, BloombergNews, WSJTax} for the importance of HFT order cancellation in the public debate.}.

The rest of the paper is organized as follows. Section \ref{Section:Model} describes the model. In Section \ref{Section:Results}, we present and discuss the simulation results. Finally, Section \ref{Section:Conclusions} concludes.

\section{The Model}
\label{Section:Model}

We model a stock market populated by heterogeneous, boundedly-rational traders.
Agents trade an asset for $T$ periods and transactions are executed through a limit-order book (LOB) where the type, the size and the price of all agents' orders are stored \footnote{See Refs. \cite{maslov2000simple, zovko2002power, farmer2005predictive, avellaneda2008high, pellizzari2009some, bartolozzi2010multi, cvitanic2010high}. For a detailed study of the statistical properties of the limit order book cf. Refs. \cite{bouchaud2002statistical, luckock2003steady, smith2003statistical}.}. Agents are classified in two groups according to their trading frequency, \textit{i.e.,} the average amount of time elapsed between two order placements. More specifically, the market is populated by $N_{L}$ low-frequency (LF) and $N_{H}$ high-frequency (HF) traders ($N=N_{L}+N_{H}$). Note that, even if the number of agents in the two groups is kept fixed over the simulations, the proportion of low- and high-frequency traders changes over time as some agents may not be active in each trading session. Moreover, agents in the two groups are different not only in terms of trading frequencies, but also in terms of strategies and activation rules. A detailed description of the behavior of LF and HF traders is provided in Sections \ref{Subsection:LFT} and \ref{Subsection:HFT}. We first present the timeline of events of a representative trading session (cf. Section \ref{Subsection:Timeline}).

\subsection{The Timeline of Events}
\label{Subsection:Timeline}

At the beginning of each trading session $t$, active LF and HF agents know the past closing price as well as the past and current fundamental values 
According to the foregoing information set, in each session $t$, trading proceeds as follows:

\begin{enumerate}
\item Active LF traders submit their buy/sell orders to the LOB market, specifying their size and limit price.
\item Knowing the orders of LF traders, active HF agents start trading \textit{sequentially} and submit their buy/sell orders. The size and the price of their orders are also listed in the LOB.
\item LF and HF agents' orders are matched and executed\footnote{The price of an executed contract is the average between the matched bid and ask quotes.} according to their price and then arrival time. Unexecuted orders rest in the LOB for the next trading session.
\item At the end of the trading session, the closing price ($\bar{P}_{t}$) is determined. The closing price is the maximum price of all executed transactions in the session. 
\item Given $\bar{P}_{t}$, all agents compute their profits and LF agents update their strategy for the next trading session (see Section \ref{Subsection:LFT} below).
\end{enumerate}

\subsection{Low-Frequency Traders\label{Subsection:LFT}} 

In the market, there are $i=1,\dots,N_{L}$ low-frequency agents who take short or long positions on the traded asset. The trading frequency of LF agents is based on \textit{chronological} time, \textit{i.e.} it is exogenous and constant over time. In particular, LF agents' trading speed is drawn from a truncated exponential distribution with mean $\theta$ and bounded between $\theta_{min}$ and $\theta_{max}$ minutes.

In line with most heterogeneous-agent models of financial markets, LF agents determine the quantities bought or sold (\textit{i.e.,} their orders) according to either a fundamentalist or a chartist (trend-following) strategy \footnote{See, e.g., Refs.  \cite{chiarella1992dynamics, lux1995herd, lux2000volatility, farmer2002market, chiarella2003heterogeneous, hommes2005robust, chiarella2006dynamic, westerhoff2008use}.}. More precisely, given the last closing price $\bar{P}_{t-1}$, orders under the chartist strategy ($D^{c}_{i,t}$) are determined as follows:
\begin{equation}
D^{c}_{i,t}=\alpha^{c}(\bar{P}_{t-1}-\bar{P}_{t-2})+ \epsilon^{c}_{t},
\end{equation}
where $0 < \alpha^{c} < 1$ and $\epsilon^{c}_{t}$ is an i.i.d. Gaussian stochastic variable with zero mean and $\sigma^{c}$ standard deviation. If a LF agent follows a fundamentalist strategy, her orders ($D^{f}_{i,t}$) are equal to: 
\begin{equation}
D^{f}_{i,t}=\alpha^{f}(F_{t}-\bar{P}_{t-1})+ \epsilon^{f}_{t},
\end{equation}
where $0 < \alpha^{f} < 1$ and $\epsilon^{f}_{t}$ is an i.i.d. normal random variable with zero mean and $\sigma^{f}$ standard deviation. The fundamental value of the asset $F_t$ evolves according to a geometric random walk:
\begin{equation}
F_{t}=F_{t-1}(1+\delta)(1+y_{t}),
\end{equation}
with i.i.d. $y_{t} \sim N(0,\sigma^{y})$ and a constant term $\delta > 0$. After $\gamma^{L}$ periods, unexecuted orders expire, \textit{i.e.} they are automatically withdrawn from the LOB. Finally, the limit-order price of each LF trader is determined by:
\begin{equation}
P_{i,t}=\bar{P}_{t-1}(1+\delta)(1+z_{i,t}),
\end{equation}
where $z_{i,t}$ measures the number of ticks away from the last closing price $\bar{P}_{t-1}$ and it is drawn from a Gaussian distribution with zero mean and $\sigma^{z}$ standard deviation. 

In each period, low-frequency traders can switch their strategies according to their profitability. At the end of each trading session $t$, once the closing price $\bar{P}_{t}$ is determined, LF agent $i$ computes her profits ($\pi^{st}_{i,t}$) under chartist ($st=c$) and fundamentalist ($st=f$) trading strategies as follows:
\begin{equation}
\pi^{st}_{i,t}=(\bar{P}_{t}-P_{i,t})D^{st}_{i,t}.
\end{equation}
Following Refs. \cite{brock1998heterogeneous,westerhoff2008use,pellizzari2009some}, the probability that a LF trader will follow a chartist rule in the next period ($\Phi^{c}_{i,t}$) is given by:
\begin{equation}
\Phi^{c}_{i,t}=\frac{e^{\pi^{c}_{i,t}/\zeta}}{e^{\pi^{c}_{i,t}/{\zeta}} + e^{\pi^{f}_{i,t}/{\zeta}}},
\end{equation}
with a positive intensity of switching parameter $\zeta$. Accordingly, the probability that LF agent $i$ will use a fundamentalist strategy is equal to $\Phi^{f}_{i,t}=1-\Phi^{c}_{i,t}$.

\subsection{High-Frequency Traders\label{Subsection:HFT}} 

As mentioned above, the market is also populated by $j=1,\dots, N_{H}$ high-frequency agents who buy and sell the asset  \footnote{We assume that $N_{H}<N_{L}$. The proportion of HF agents vis-\`{a}-vis LF ones is in line with empirical evidence \cite{kirilenko2011flash,paddrik2011}.}.

HF agents differ from LF ones not only in terms of trade speed, but also in terms of activation and trading rules. In particular, contrary to LF strategies, which are based on chronological time, the algorithmic trading underlying the implementation of HFT naturally leads HF agents to adopt trading rules  framed in \textit{event} time \cite{easley2012volume} \footnote{On the case for moving away from chronological time in modeling financial series see Refs. \cite{MnT67,cla73,AnG00}.}. More precisely, we assume that the activation of HF agents depends on the extent of price fluctuations observed in the market. As a consequence, HF agents' trading speed is \textit{endogenous}. Each HF trader has a fixed price threshold $\Delta x_{j}$, drawn from a uniform distribution with support bounded between $\eta_{min}$ and $\eta_{max}$. This determines whether she will participate or not in the trading session $t$ \footnote{See Ref. \cite{aloud2012modelling} for a
similar attempt in this direction.}:
\begin{equation}
\Bigg| \frac{\bar{P}_{t-1}-\bar{P}_{t-2}}{\bar{P}_{t-2}} \Bigg| > \Delta x_{j}. \label{eq:HFTactivation}
\end{equation}
Active HF agents submit buy or sell limit orders with equal probability $p=0.5$ \cite{maslov2000simple, farmer2005predictive}.

Furthermore, HF traders adopt \textit{directional} strategies that try to profit from the anticipation of price movements \cite{securities2010equity,aloud2011directional}. To do this, HF agents exploit the price and order information released by LF agents. 

First, HF traders determine their buy (\textit{sell}) order size, $D_{j,t}$, according to the volumes available in the opposite side of the LOB. More specifically, HF traders' order size is drawn from a truncated Poisson distribution whose mean depends on volumes available in the sell(\textit{buy})-side of the LOB if the order is a buy (\textit{sell}) order \footnote{In the computation of this mean, the relevant market volumes are weighted by the parameter $0 < \lambda < 1$. This link between market volumes and HF traders' order size is motivated by the empirical evidence which suggests that HF traders typically submit large orders \cite{kirilenko2011flash}. Moreover, empirical works also indicate that HF traders do not accumulate large net positions. Thus, we introduce two additional constraints to HF order size. First, HF traders' net position is bounded between +/-3,000. Second, HF traders' buy (sell) orders are smaller than one quarter of the total volume present in the sell (buy) side of the LOB \cite{kirilenko2011flash,bartolozzi2010multi, paddrik2011}.}.  As HF traders adjust the volumes of their orders to the ones available in the LOB, they manage to absorb LF agents' orders.

Second, in each trading session $t$, HF agents trade near the best ask ($P^{ask}_t$) and bid ($P^{bid}_t$) prices available in the LOB \cite{paddrik2011}. This assumption is consistent with empirical evidence on HF agents' behavior which suggest that most of their orders are placed very close to the last best prices \cite{securities2010findings}. Accordingly, HF buyers and sellers' limit prices are formed as follows:
\begin{equation} \label{eq:HFTprice}
P_{j,t}=P^{ask}_{t}(1+\kappa_{j}) \ \ \ \ \ \ \ \ \ \ \ \ \ \ \
P_{j,t}=P^{bid}_{t}(1-\kappa_{j}),
\end{equation}
where $\kappa_{j}$ is drawn from a uniform distribution with support ($\kappa_{min},\kappa_{max}$).

A key characteristic of empirically-observed high-frequency trading is the high order cancellation rate \cite{securities2010findings, kirilenko2011flash}. We introduce such a feature in the model by assuming that HF agents' unexecuted orders are automatically removed from the LOB after a period of time $\gamma^H$, which is shorter than LF agents' one, \textit{i.e.} $\gamma^H < \gamma^{L}$. Finally, at the end of each trading session, HF traders' profits ($\pi_{j,t}$) are computed as follows \footnote{Simulation exercises in the baseline scenario reveal that the strategies adopted by HF traders are able to generate positive profits, thus justifying their adoption by HF agents in the model. In particular, simulation results reveal that the distribution of HF traders' profits is skewed to the right and has a positive mean.}:
\begin{equation}
\pi_{j,t}=(\bar{P}_{t}-P_{j,t})D_{j,t}.
\end{equation} 
where $D_{j,t}$ is the HF agent's order size, $P_{j,t}$ is her limit price and $\bar{P}_{t}$ is the market closing price.

\section{Simulation Results}\label{Section:Results}

\noindent We investigate the properties of the model presented in the
previous section via extensive Monte-Carlo simulations. More
precisely, we carry out $MC=50$ Monte-Carlo iterations, each one 
composed of $T=1,200$ trading sessions using the baseline parametrization,
described in Table \ref{tab:paramValues}. As a first step in our analysis of
simulation results, we check whether the model is able to account for the main stylized facts of financial markets (see Section \ref{sec:high-freq-trad}). We then assess whether the model can generate flash crashes characterized by empirical properties close to the ones observed in real data (cf. Section \ref{sec:flash}) and we investigate the determinants of flash crashes (cf. Section \ref{sec:anat-flash-crash}). Finally, we study post flash-crash recoveries by investigating the consequences of different degrees of HF traders' order cancellation on model dynamics (see Section \ref{sec:order-canc-flash}).

\subsection{Stylized Facts of Financial Markets}
\label{sec:high-freq-trad}

\begin{figure}[tbp]
  \centering
  \includegraphics[width=9cm]{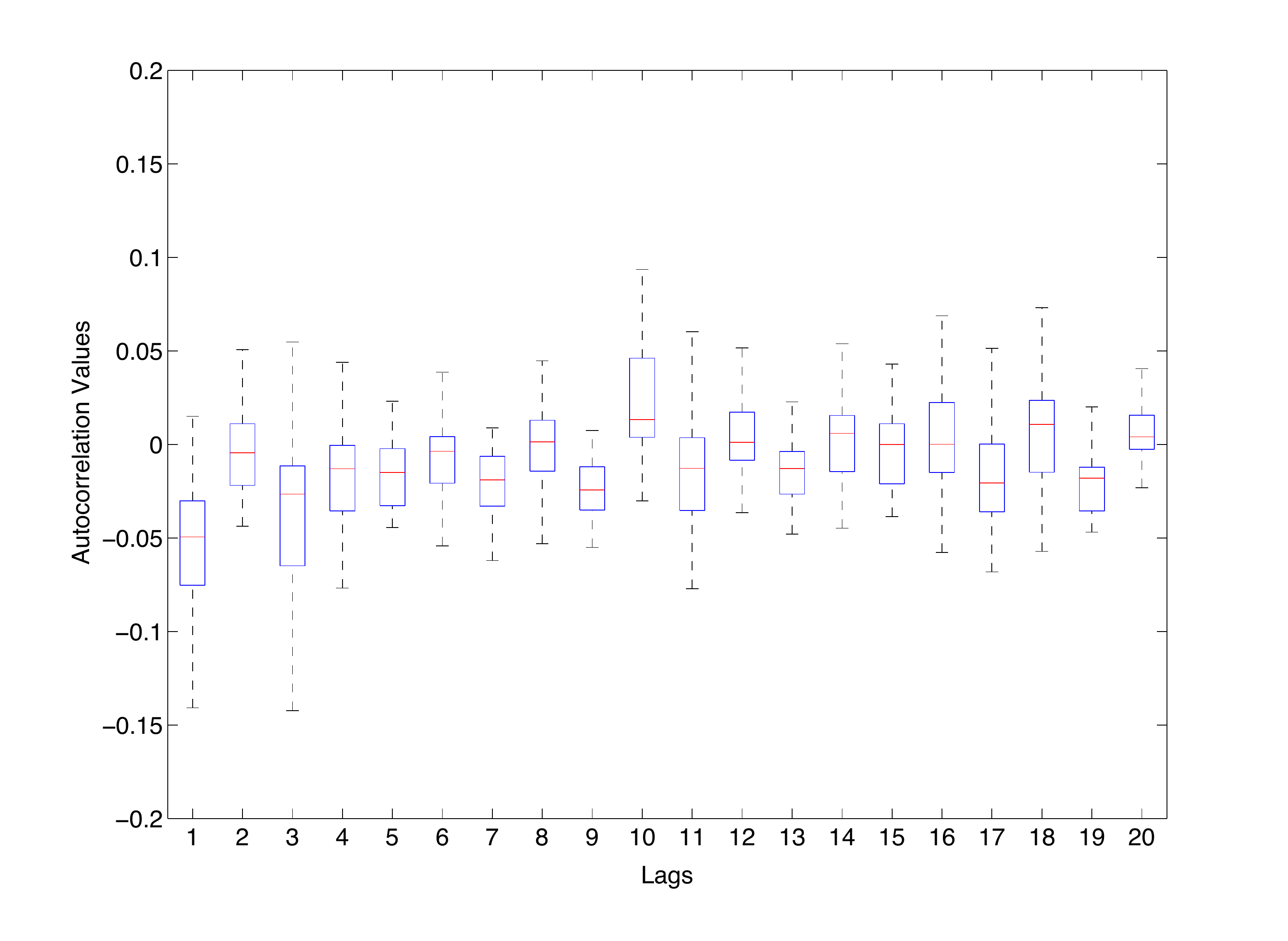}
  \caption{Box-Whisker plots of price-returns autocorrelations. Each
    plot relates to auto-correlation values for a given lag
    across independent 50 Monte-Carlo runs.}
  \label{fig:pricereturns_autocorrelations}   
\end{figure}

\begin{figure}[tbp] 
\centering   
 \includegraphics[width=9cm]{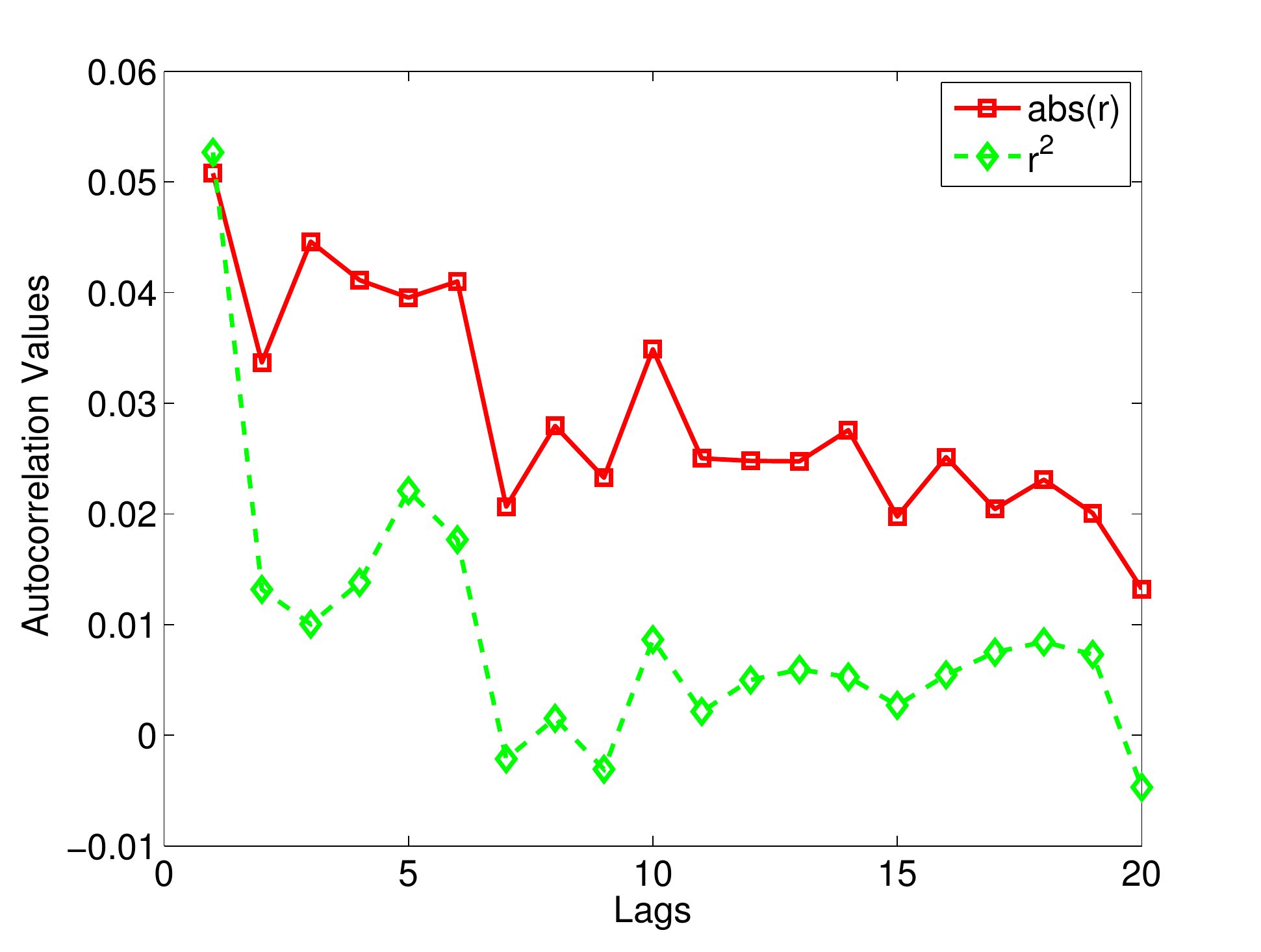}
   \caption{Sample autocorrelation functions of absolute price returns (solid line) and squared price returns (dashed line). Values are averages
    across independent 50 Monte-Carlo runs.}
  \label{fig:volatilityclusters_autocorrelations}   
\end{figure}

\begin{figure}[tbp]
  \centering
  \includegraphics[width=9cm]{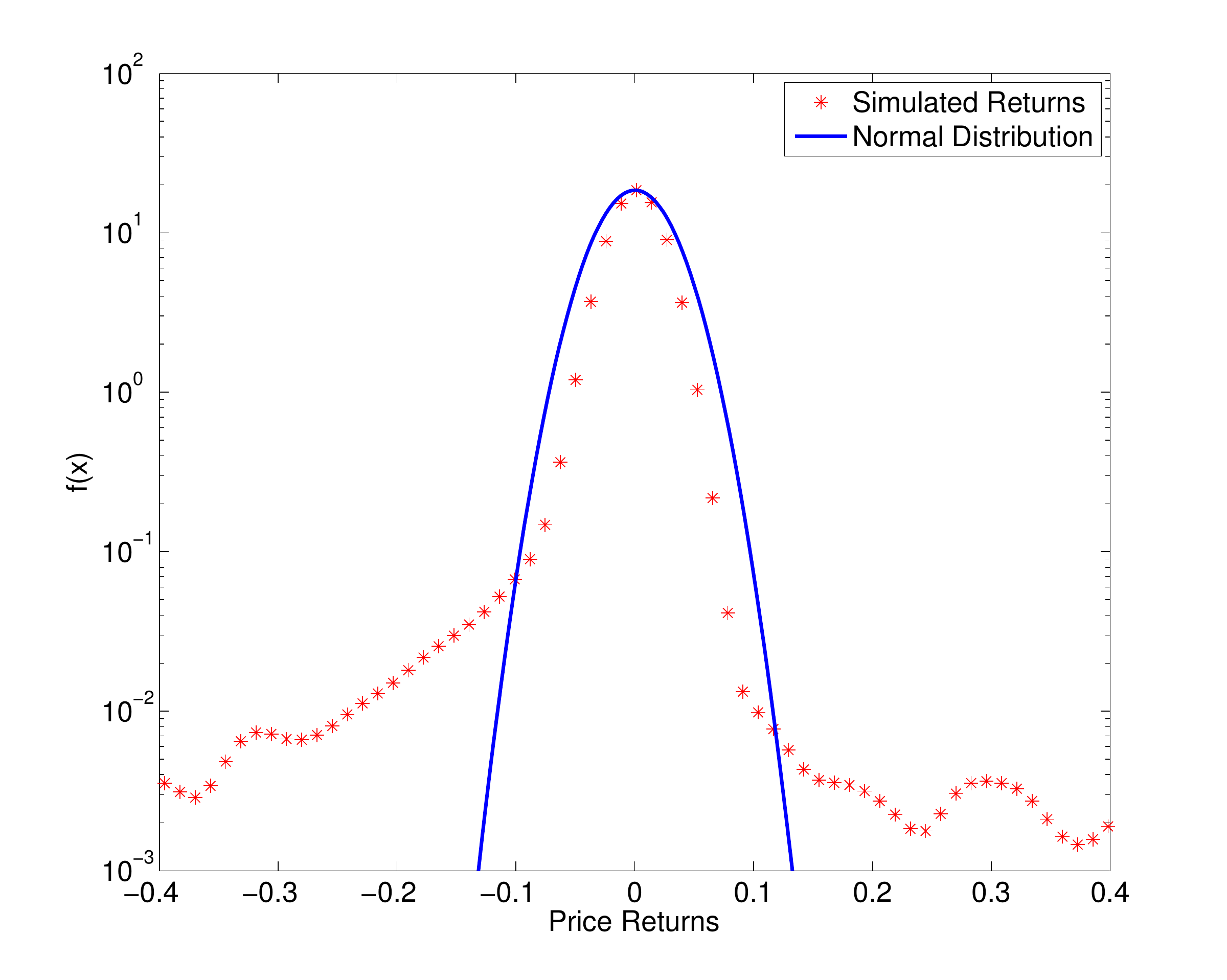}
  \caption{Density of pooled price returns (stars) across 50 independent
    Monte-Carlo runs together with a Normal fit (solid
    line). Logarithmic scale on y-axis. Densities are estimated using
    a kernel density estimator using a bandwidth optimized for Normal
    distributions.}
\label{fig:returns_density}
\end{figure}

\begin{figure}[tbp]
 \centering   
 \includegraphics[width=9cm]{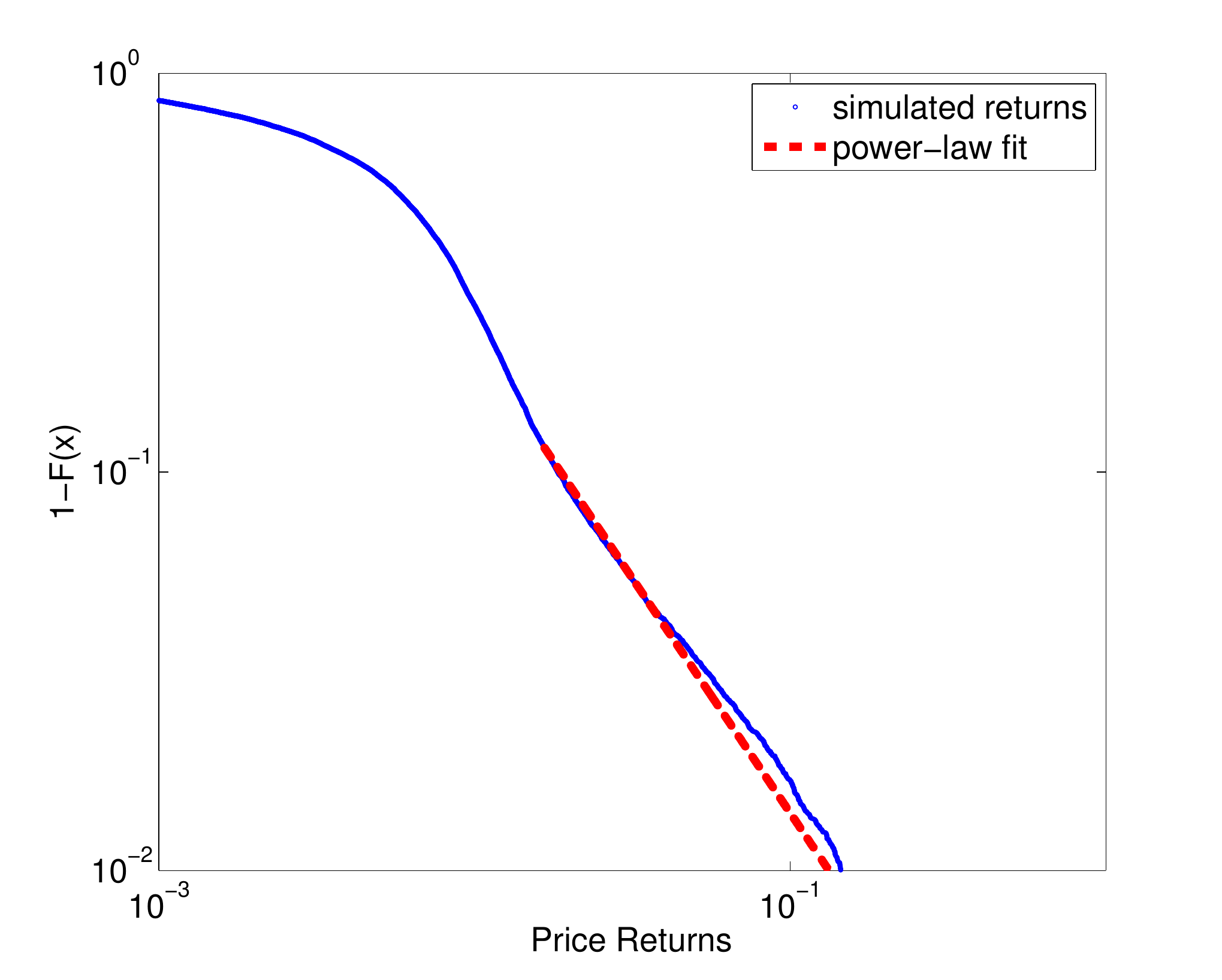}
 \caption{Complementary cumulative distribution of negative
   price returns (circles) together with power-law
   fit (dashed line). Double-logarithmic scale.}
\label{fig:power_law_returns}
\end{figure}

How does the model fare in terms of its ability to replicate the main
statistical properties of financial markets? First, in line with the empirical evidence of zero autocorrelation detected in price returns \cite{f70, pagan1996econometrics, chakraborti2011econophysics}, we find that model-generated autocorrelation values of price-returns (calculated as logarithmic differences) do not reveal any significant pattern and are always not significantly different from zero (see the box-whisker plots in Figure \ref{fig:pricereturns_autocorrelations}) \footnote{More precisely, the confidence interval of the values at each lag (measured by the extent of the whiskers) always includes the zero. Notice that each whiskers' length in the plot corresponds to 99.3\% data coverage under the assumption that autocorrelation values are drawn from a Normal distribution.}. Moreover, in contrast to price returns, the autocorrelation functions of both absolute and squared returns display a slow decaying pattern, which is more pronounced in the case of absolute returns (cf. Figure \ref{fig:volatilityclusters_autocorrelations}). This indicates the presence of volatility clustering in our simulated data \cite{mandelbrot1963variation, cont1997scaling, lo1999non}. Note also that such autocorrelation values are very similar to empirically-observed ones \cite{cont2001empirical}.

Another robust statistical property of financial markets is the existence of fat tails in the distribution of price returns. To investigate the presence of such a property in our simulated data, we plot in Figure \ref{fig:returns_density} the density of pooled returns across Monte-Carlo runs (stars) together with a normal density (solid line) fitted on the pooled sample. As the figure shows quite starkly, the distribution of price returns significantly departs from the Gaussian benchmark \cite{mandelbrot1963variation, cont2001empirical}. The departure from normality is particularly evident in the tails (see Figure \ref{fig:power_law_returns}), which are well approximated by a power-law density \cite{lux2006financial}.

\begingroup
     \squeezetable
\begin{table}[tbp]
\caption{Volatility and flash-crash statistics across different scenarios. Values are averages across 50 independent Monte-Carlo runs. Monte-Carlo standard errors in parentheses.}
  \centering
  \begin{tabular}{lccc}
\\
\hline \hline \noalign{\smallskip}
          Scenario&$\sigma_P$&Number of& Avg. duration of \\
                         &                               & flash crashes &
    flash crashes \\  
\hline              
Baseline scenario & 0.020 & 8.800 & 14.069 \\
 & (0.001) & (0.578) & (0.430) \\
Only-LFT scenario & 0.005 & - & - \\
 &(0.001) & - & - \\
 \noalign{\smallskip} \hline \hline 
\end{tabular}
\label{tab:volandflash_across_scenarios}
\end{table}
\endgroup

\subsection{Stylized Facts of Flash Crashes}
\label{sec:flash}

Our model appears to be quite successful in replicating the ``standard'' stylized facts about financial markets. However, is it able to account for the emergence of flash crashes \cite{securities2010findings, kirilenko2011flash}? Simulations results provide a positive answer. In line with the evidence presented in Ref. \cite{kirilenko2011flash} about the flash crash of May $6^{th}$, 2010, we identify flash crashes as drops in the asset price of at least 5\% followed by a sudden recovery of 30 minutes at maximum (corresponding to thirty trading sessions in each simulation run). Applying such a definition, we find that the model is able to endogenously generate flash crashes and their frequency is significantly higher than one (see the third column of Table \ref{tab:volandflash_across_scenarios}).   

Furthermore, the model is also able to account for another relevant and recent stylized fact observed during flash crashes, namely the negative correlation between price and volumes \cite{kirilenko2011flash}. To illustrate this, we report in Table \ref{tab:price_corr} unconditional correlations between price returns and total volumes as well as correlation values conditioned on two distinct market phases: ``flash crashes'' (which includes price crashes and subsequent recoveries) and ``normal times'' (where we excluded all observations regarding flash crashes). Normal times are characterized by a weak but positive correlation between returns and volumes, which is also reflected in the low value of the unconditional correlations. In contrast, correlations turns out to be negative and much more significant during flash crashes. HF traders do appear to be responsible for this result: HFT volume displays a significant negative correlation with returns during flash crashes, whereas LFT volumes are always positively correlated with price returns (see the second and third columns of Table \ref{tab:price_corr}).

\begin{table}[tbp]
  \centering
\caption{Correlation values between price returns and different types of orders volumes. Values are averages across 50 independent Monte-Carlo runs. Monte-Carlo standard errors in parentheses.}
\label{tab:price_corr}
\begin{tabular}{lccc}
\\
\hline \hline \noalign{\smallskip}
 & Total volume & HFT volume & LFT volume \\
\hline
Unconditional & 0.018 & 0.012 & 0.096 \\
 & (0.007) & (0.007) & (0.005) \\
Flash crashes & -0.112 & -0.113 & 0.030 \\
 & (0.028) & (0.028) & (0.031) \\
Normal times & 0.023 & 0.016 & 0.103 \\
 & (0.006) & (0.006) & (0.005) \\
\noalign{\smallskip} \hline \hline 
\end{tabular}  
\end{table}

The above findings suggest a significant role of HF traders during flash crashes. However, how relevant is high-frequency trading for the emergence of the aforementioned stylized facts, and more generally of flash crashes? To check this, we carry out  a Monte-Carlo exercise in a scenario wherein only low-frequency traders are present (``only-LFT'' scenario). The comparison between such a scenario and the baseline one reveals that the main stylized facts observed in financial markets are reproduced also when we remove HF agents \footnote{The plots and statistics concerning the analysis of stylized facts in the ``only-LFT'' scenario are available from the authors upon request.}. In contrast, when the market is populated only by LF traders, flash crashes do not emerge (cf. Table \ref{tab:volandflash_across_scenarios}). Moreover, price returns volatility significantly drops providing further evidence on the destabilizing role of high-frequency trading.

To sum up, our results confirm that the model is able to reproduce the main stylized facts of financial markets. Furthermore, they indicate that the emergence of flash crashes is strongly related to the presence of HF traders in the market. In the next section, we further spotlight flash crashes, studying which features of high-frequency trading are more responsible for their emergence.

\subsection{The Anatomy of Flash Crashes}
\label{sec:anat-flash-crash}
\begin{figure}[tbp]
  \centering
  \includegraphics[width=9cm]{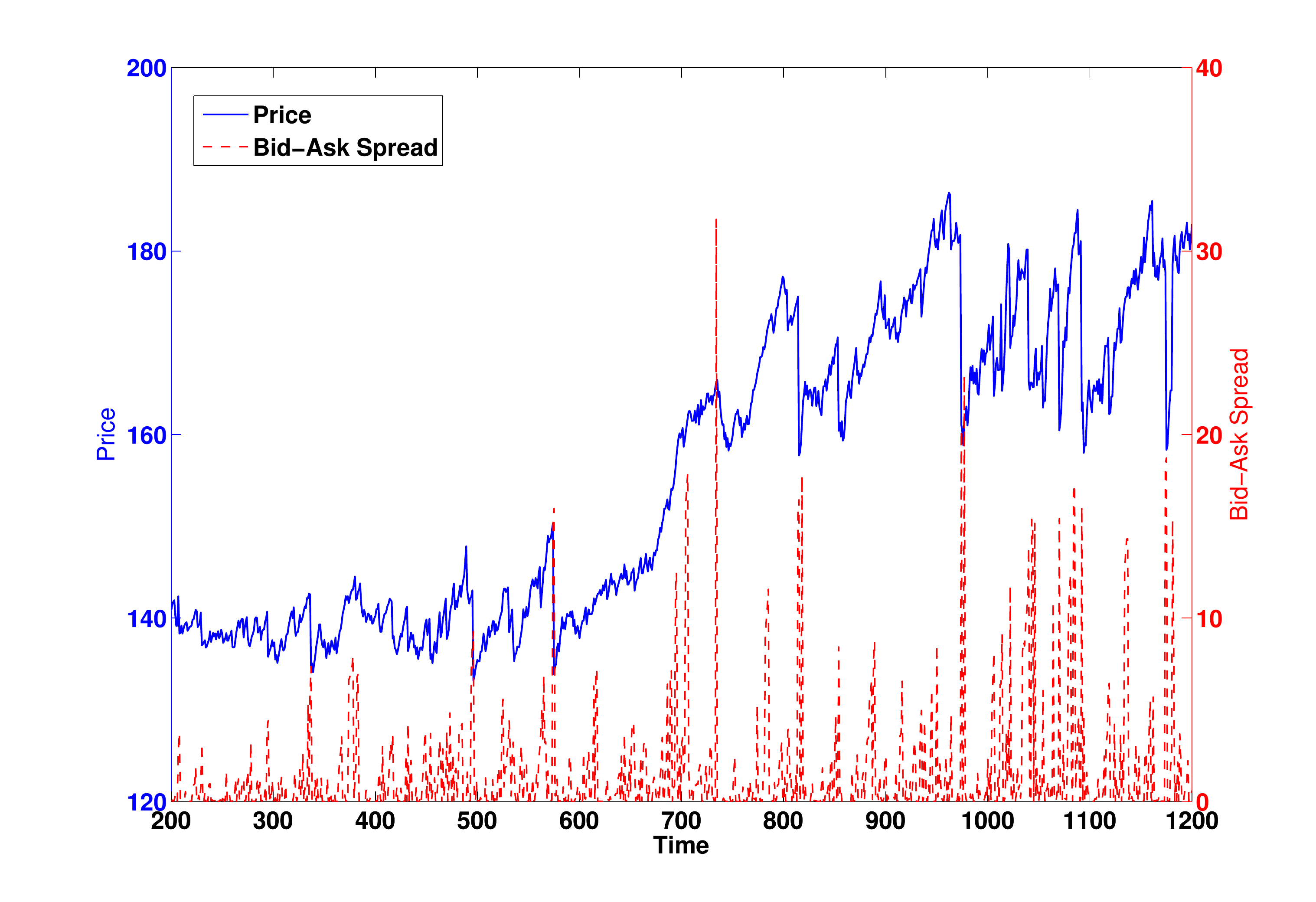}
  \caption{Evolution of asset price (solid line) and bid-ask spread (dashed line) in a single Monte-Carlo run.}
  \label{fig:price_bidaskspread}
\end{figure}

Let us begin considering the evolution of the asset price and bid-ask spread in a single simulation run (cf. Figure \ref{fig:price_bidaskspread}).~The plot reveals that sharp drops in the asset price tend to be associated with periods of large bid-ask spreads (see Ref.~\cite{doyne2004really} for empirical evidence on the relation between liquidity fluctuations and large price changes).~This piece of evidence suggests that flash crashes emerge when market liquidity is very low. 

\begin{figure}[tbp]
  \centering
  \includegraphics[width=9cm]{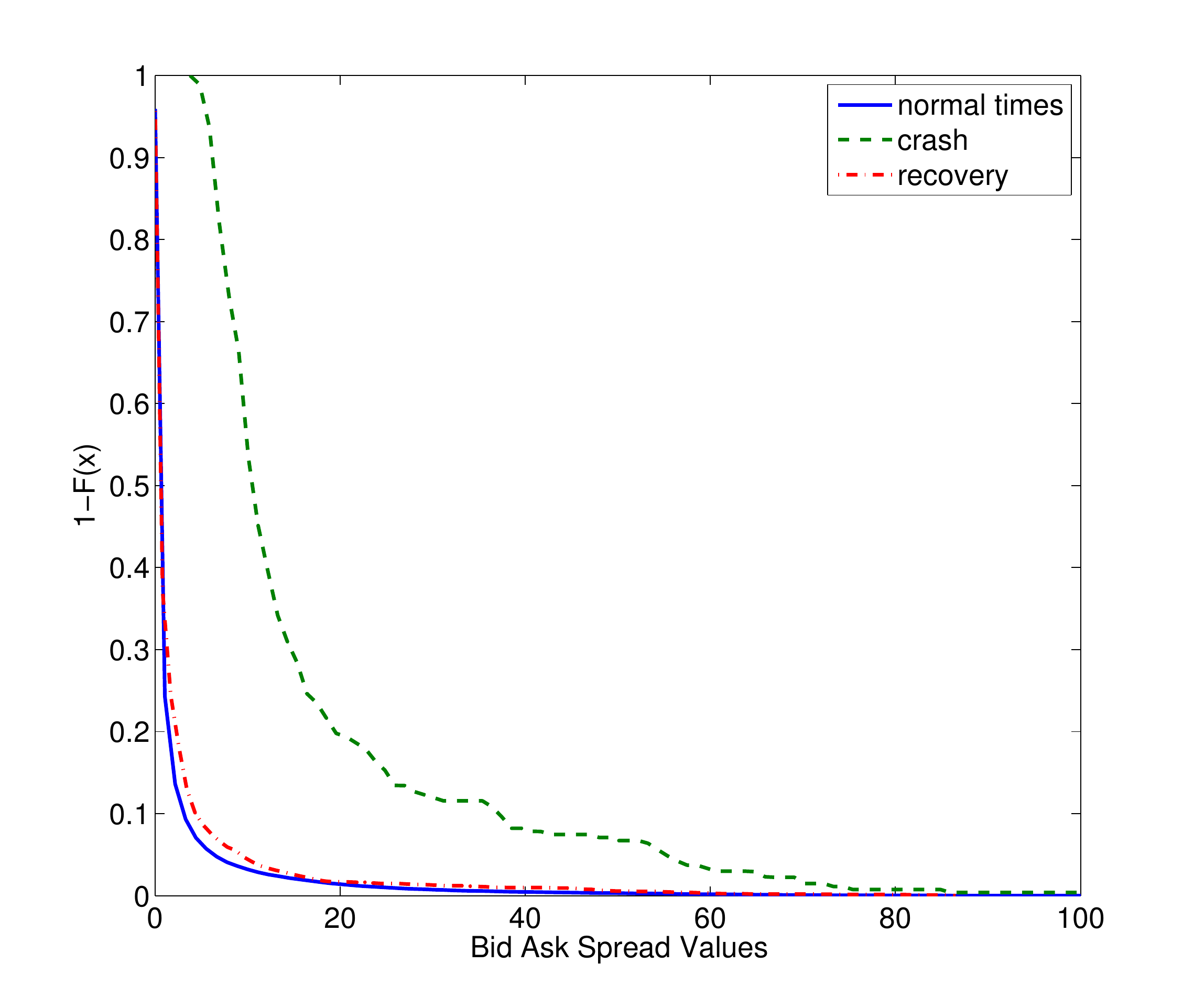}
\caption{Complementary cumulative distributions of bid-ask spreads in different market phases. Pooled sample from 50 independent Monte-Carlo runs.}
\label{fig:bidaskspread_distribution}
\end{figure}

To shed more light on the relation between flash crashes and market liquidity, we compute the distributions of bid-ask spreads conditioned on different market phases. More precisely, we construct the pooled samples (across Monte-Carlo runs) of bid-ask spread values singling out ``normal times'' phases (see also Section \ref{sec:flash}) and decomposing ``flash-crash'' periods in ``crash'' phases (\textit{i.e.} periods of sharp drops in the asset price) and the subsequent ``recovery'' phases (periods when the price grows back to its pre-crisis level). Next, we estimate the complementary cumulative distributions of bid-ask spreads in each market phase using a kernel-density estimator. The distributions plotted in Figure \ref{fig:bidaskspread_distribution} confirm what we have learned from the visual inspection of price and bid-ask spread dynamics in a single simulation run. Indeed, the mass of the distribution of bid-ask spreads is significantly shifted to the right during flash crashes vis-\`{a}-vis normal times.

The aforementioned switches between periods of high and low market liquidity
(\textit{i.e.,} between periods of low and high bid-ask spreads) are explained by the different strategies employed by high- and low-frequency traders in our model. Active LF traders set their order prices ``around'' the closing price of the last trading session. This behavior tends to fill the gap existing between the best bid and ask prices at the beginning of a given trading section. Instead, active HF traders send their orders after LF agents and place large buy (\textit{sell}) orders just few ticks above (\textit{below}) the best ask (\textit{bid}). Such a pricing behavior widens the difference between bid and ask prices in the LOB. As a consequence, the directional strategies of HF traders can lead to wide bid-ask spreads, setting the premises for the emergence of flash crashes. However, large spreads are not enough to generate significant drops in the closing price if LF and HF agents' orders are evenly distributed in the LOB between the buy and sell sides. Accordingly, as the closing price is the maximum price of all executed transactions in the book, an even distribution of orders should lead to small fluctuations in the closing price. In contrast, extreme price fluctuations require concentration of orders on one side of the book.

\begin{figure}[tbp]
  \centering
  \includegraphics[width=9cm]{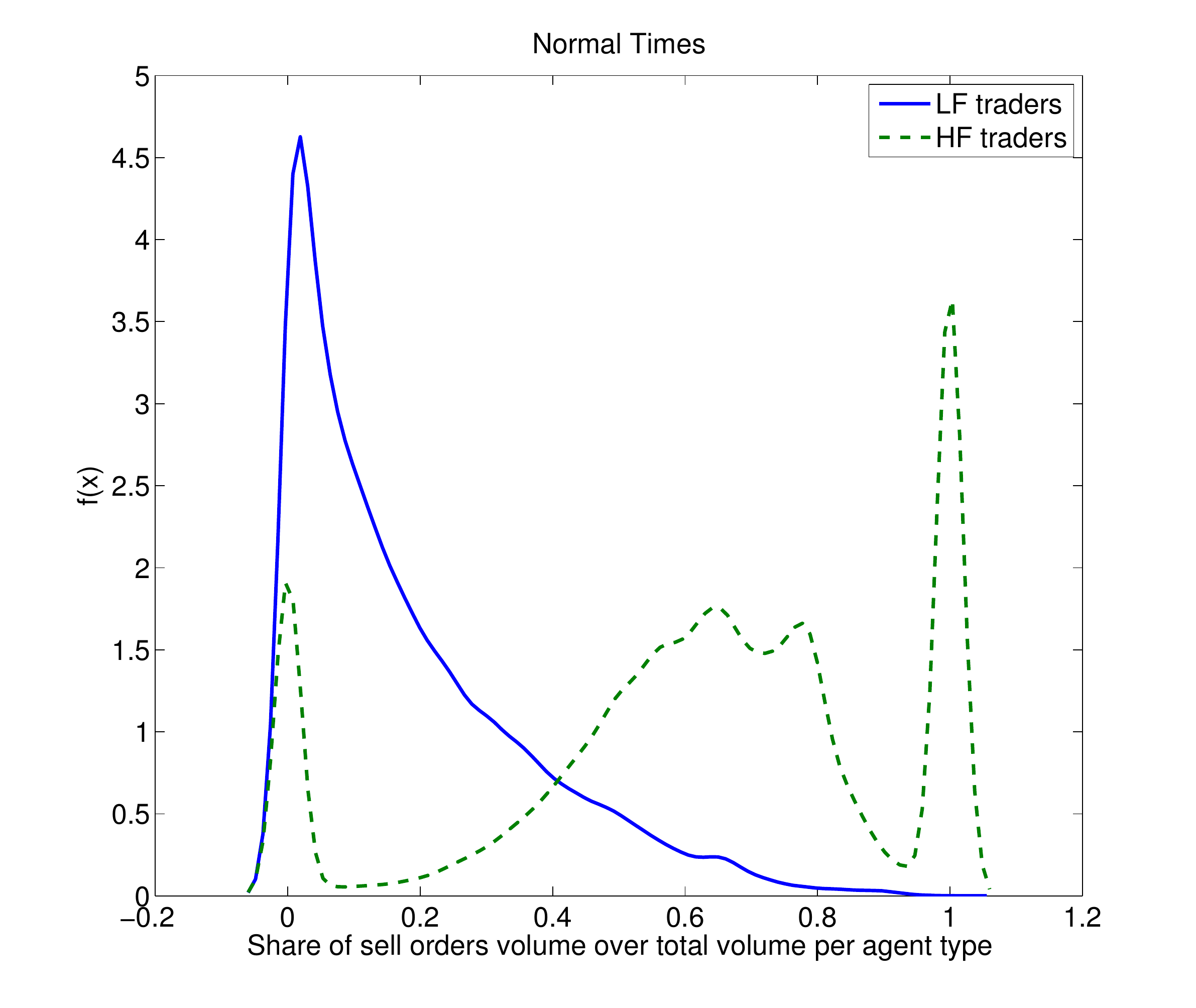}
  \caption{Kernel densities of shares of sell order volume over total order
  volume of the same agent type (HFTs: solid line, LFTs: dashed
  line). Normal times.}
\label{fig:concentration_ratios_HFTvsLFT_nt}
\end{figure}

\begin{figure}[tbp]
  \centering
  \includegraphics[width=9cm]{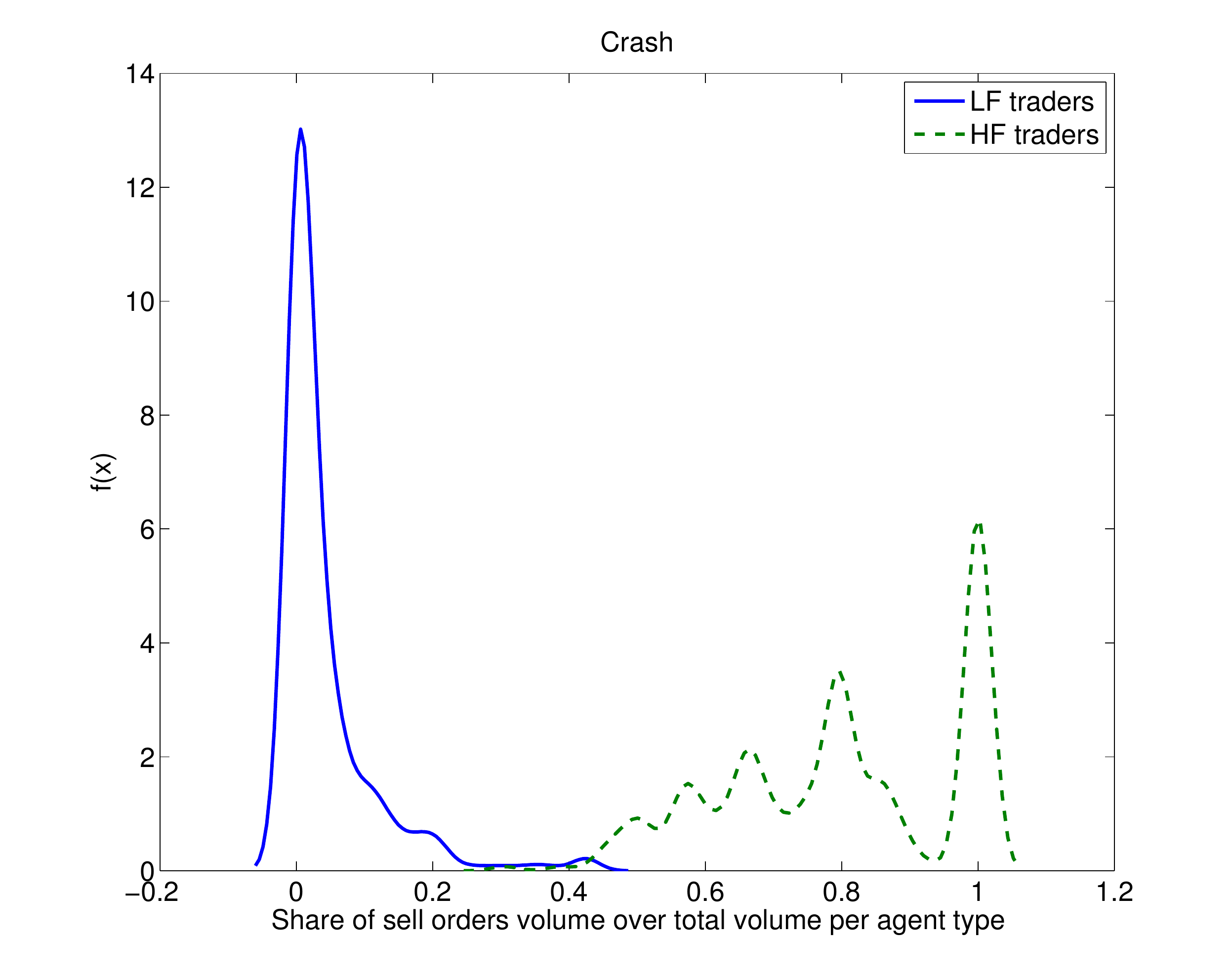}
  \caption{Kernel densities of shares of sell order volume over total order
  volume of the same agent type (HFTs: solid line, LFTs: dashed
  line). Flash Crashes.}
\label{fig:concentration_ratios_HFTvsLFT_flash}
\end{figure}

To further explore such conjecture, we analyze the distributions of shares of sell order volumes in the book made by each type of agent (HF or LF traders) over the total volume within the same category. This ratio captures the concentration of the orders on the sell side of the LOB disaggregated for agents' type. In particular, the more the sell concentration ratio is close to one, the more a given category of agents (\textit{e.g.,} HF traders) is filling the LOB with sell orders. Figures \ref{fig:concentration_ratios_HFTvsLFT_nt} and \ref{fig:concentration_ratios_HFTvsLFT_flash} compare kernel densities of the foregoing sell concentration ratios for HF and LF agents in normal times
and crashes, respectively. Let us start examining the latter. First, Figure \ref{fig:concentration_ratios_HFTvsLFT_flash} shows that during crashes the supports of the LF and HF traders' distributions do not overlap. This hints to a very different behavior of LF and HF agents during flash crashes. Second, during crash times, LF and HF traders' orders are concentrated on opposite sides of the LOB. More specifically, the mass of the distribution of LF agents' orders is concentrated on the buy side of the book, whereas the mass of the HF traders' distribution is found on very high values of the sell concentration ratio (see Figure \ref{fig:concentration_ratios_HFTvsLFT_flash}). These extreme behaviors are not observed during normal times
(cf. Figure \ref{fig:concentration_ratios_HFTvsLFT_nt}). Indeed, in tranquil market phases, the supports of the LFT and HFT densities overlap and they encompass the whole support of the sell concentration statistic. 

\begin{figure}[tbp]
  \centering
  \includegraphics[width=9cm]{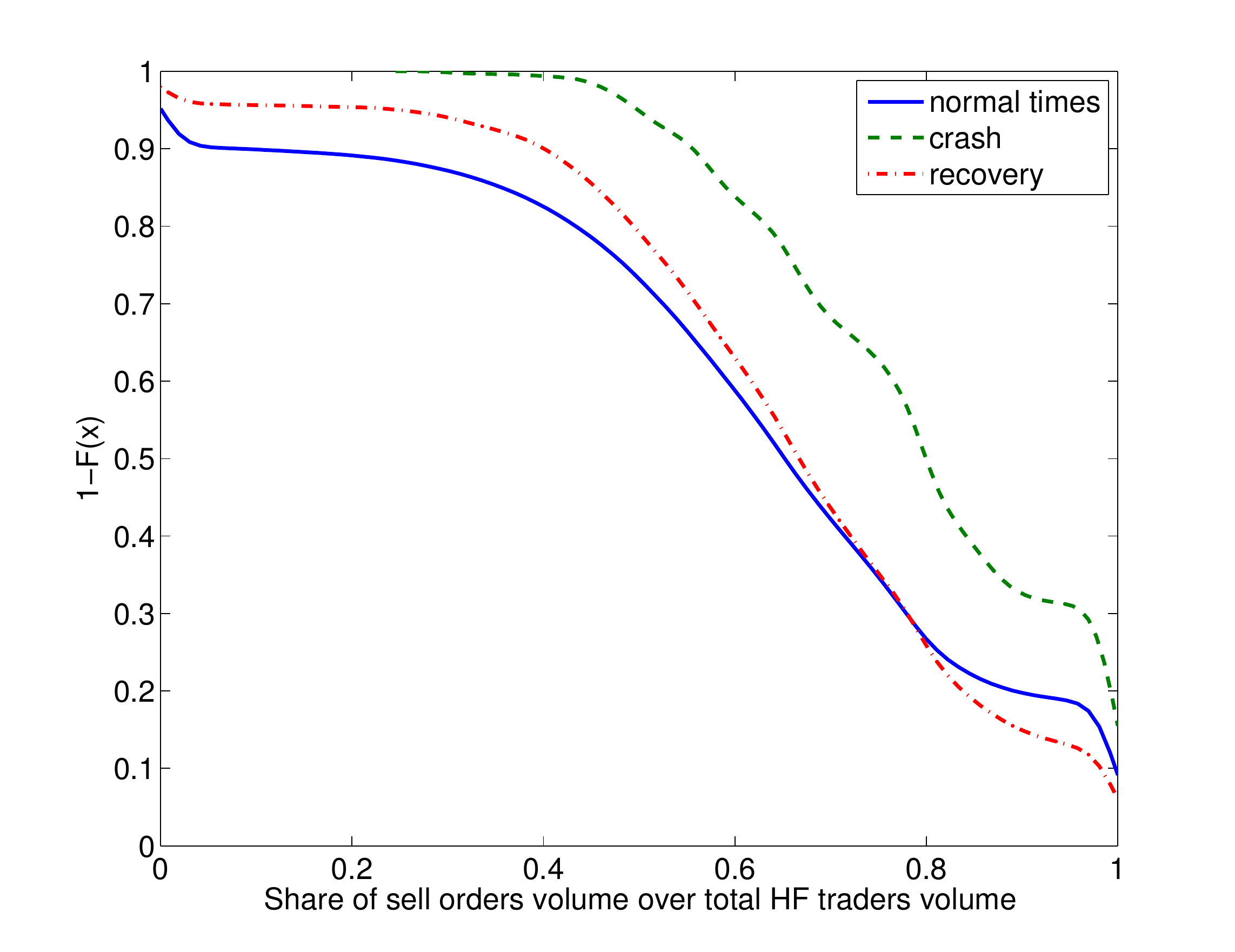}
  \caption{Complementary cumulative distribution of shares of sell order volume over total orders volume of the same agent type for different market
  phases. HFT orders.}
\label{fig:concentration_ratios_cumulative_HFT}
\end{figure}
\begin{figure}[tbp]
  \centering
  \includegraphics[width=9cm]{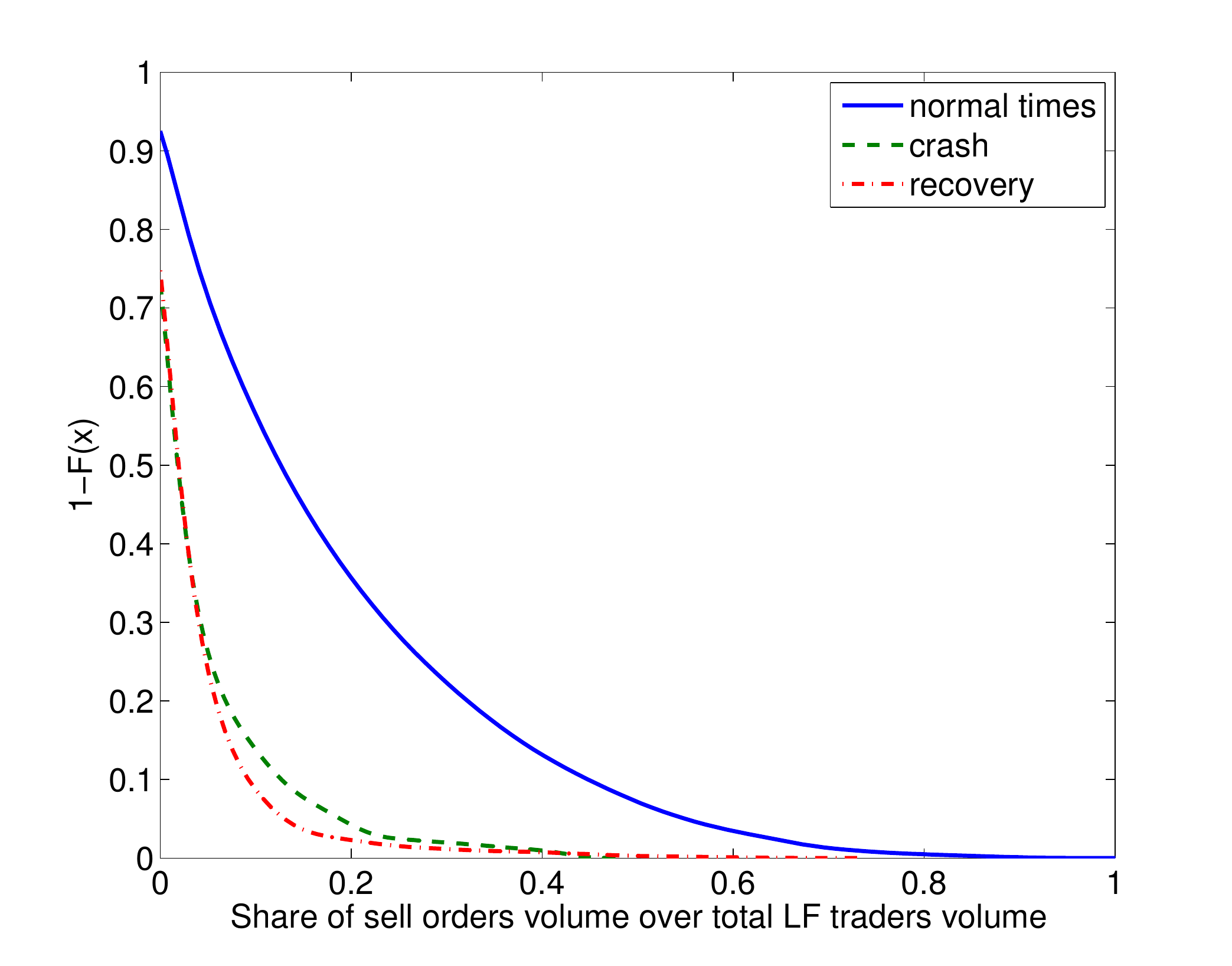}
  \caption{Complementary cumulative distribution of shares of sell order volume over total order volume of the same agent type for different market
  phases. LFT orders.}
\label{fig:concentration_ratios_cumulative_LFT}
\end{figure}

We further study the above differences in order behavior, analyzing the complementary cumulative distributions of the sell concentration statistic for the same type of agent and across different market phases (cf. Figures \ref{fig:concentration_ratios_cumulative_HFT} and \ref{fig:concentration_ratios_cumulative_LFT}). The complementary cumulative distributions  confirm that flash crashes are generated by the concentration of HF and LF orders on opposite sides of the LOB. Indeed, during flash crashes, the distribution of HF orders significantly shifts to the sell side of LOB, whereas the one of LF orders moves to the left, revealing a strong concentration on buy orders.

The above discussion shows that flash crashes are a true emergent property of the model generated by the joint occurrence of three distinct events: i) the presence of a large bid-ask spread; ii) a strong concentration of HF traders' orders on the sell side of the LOB; iii) a strong concentration of LF traders' orders on the buy side of the LOB. In particular, the first two elements are in line with the empirical evidence about the market dynamics observed during the flash crash of May $6^{th}$, 2010  \cite{securities2010findings,kirilenko2011flash} and confirm the key role played by high-frequency trading in generating such extreme events in financial markets. Indeed, the emergence of periods of high market illiquidity is intimately related to the pricing strategies of HF traders (see Eq. \ref{eq:HFTprice}). Moreover, and in line with previous agent-based models in the literature
\cite{brock1998heterogeneous, westerhoff2008use, pellizzari2009some}, the synchronization of LF traders on the buy side of the LOB can be explained on the grounds of profitability-based switching behavior by such type of agents (see Section \ref{Subsection:LFT}). 

The concentration of HF traders' orders on the sell side of the book is at first glance more puzzling, given that the choice of each HF agent between selling or buying is a Bernoulli distributed variable with probability $p=0.5$. However, the spontaneous synchronization of orders becomes possible once we consider that HF agents adopt \textit{event}-time trading strategies, which lead to the emergence of price-dependent activation processes (cf. Eq. \ref{eq:HFTactivation}). Indeed, the fact that the type of order choice is Bernoulli-distributed implies that the total number of sell orders placed by active HF traders in any given session is a binomially-distributed random variable dependent on the number of active HF agents. More precisely, let $n$ be the number of active HF traders at time $t$, the probability that a fraction $k$ of these agents place a sell order is:

\begin{equation*}
{n \choose nk}p^{nk}(1-p)^{n(1-k)},
\end{equation*} and it is inversely related to $n$. Hence, the endogenous activation of HF traders coupled with heterogeneous price activation thresholds can considerably shrink the sample of active HF agents in a trading session. The smaller sample size increases the probability of observing a concentration of HF agents' orders on the sell side of the LOB which can be conducive to the emergence of a flash crash.

\subsection{Accounting for Post-Crash Recoveries}
\label{sec:order-canc-flash}

A hallmark of flash crash episodes is the fast recoveries that follow the initial huge price drop. Which factors are responsible for such rapid switches in price dynamics? Figures \ref{fig:bidaskspread_distribution} and \ref{fig:concentration_ratios_cumulative_HFT} provide insightful information on the characteristics of the post flash-crash recoveries. First, the distribution of the bid-ask spreads in a recovery is not statistically different from the one observed in normal times (see Figure \ref{fig:bidaskspread_distribution}). This shows that high spreads are not persistent and the market is able to quickly restore good liquidity conditions after a crash. Moreover, the high concentration of HF traders on the sell side of the book disappears after the crash (cf. Figure \ref{fig:concentration_ratios_cumulative_HFT}). Indeed, the distribution of the concentration ratios during recoveries is not different from the one observed in normal times. 

Two particular features of the model explain the characteristics of the recovery phases depicted above. The first is the surge in the order volumes of HF agents in the aftermath of a crash. Wide variations in the asset prices indeed trigger the activation of a large number of high-frequency traders. Accordingly, their orders will tend to be equally split between the sell and buy sides of the LOB (see the discussion in Section \ref{sec:anat-flash-crash} above), as it is also shown by the leftward shift of the distribution of the sell concentration ratio during recoveries (cf. Figure \ref{fig:concentration_ratios_cumulative_HFT}). This fast increase of the order volumes of HF agents contributes to explain the quick recovery of the closing price, as now more and more contracts will be executed at prices close to \textit{both} the best bid and ask. The second element supporting the rapid price recovery is the order-cancellation rate of HF traders. In line with empirical evidence \cite{hasbrouck2009technology}, order cancellation of HF agents is very high in the baseline scenario, as all unexecuted orders are withdrawn at the end of each trading session (see also Table \ref{tab:paramValues}). Such ``extreme'' order-cancellation behavior of HF traders implies that their bid and ask quotes always reflect current market conditions (we call it \textit{order-memory effect}, which in this case is low). This explains the low time persistence of high bid-ask spreads after a crash and contributes to the quick recovery of market liquidity and price. 

\begin{table}[tbp]
  \centering
\caption{HF traders' order cancellation rates, price volatility and flash crash statistics. Values are averages across 50 independent Monte-Carlo runs. Monte-Carlo standard errors in parentheses.}
\label{tab:ordercancellation_table}
\begin{tabular}{lccc}
\hline \hline\noalign{\smallskip}
    $\gamma^H$&$\sigma_P$&Number of& Avg. duration of \\
                         &                               & flash crashes &
    flash crashes \\
\noalign{\smallskip}\hline\noalign{\smallskip}
  1 & 0.020 & 8.800 & 14.069 \\
 & (0.001) & (0.578) & (0.430) \\
5 & 0.013 & 3.095 & 17.376 \\
 & (0.001) & (0.231) & (0.729) \\
10 & 0.010 & 2.138 & 20.076 \\
 & (0.001) & (0.155) & (0.809) \\
15 & 0.009 & 1.667 & 18.952 \\
 & (0.001) & (0.137) & (1.091) \\
20 & 0.007 & 1.000 & 24.000 \\
 & (0.001) & (0.001) & (1.259) \\
\noalign{\smallskip}\hline \hline
\end{tabular}
\end{table}

The foregoing considerations point to a positive role played by fast HF traders' order cancellation in restoring good market conditions thus explaining the low duration of flash crashes. However, high order cancellation rates also indicate high aggressiveness of HF traders in exploiting the orders placed by LF agents in the LOB (we call it \textit{liquidity-fishing effect}). This favors the emergence of high bid-ask spreads in the market thus increasing the probability of observing a large fall in the asset price. 

To further explore the role of HF traders' order cancellation on price
fluctuations, we perform a Monte-Carlo experiment where we vary
the number of periods an unexecuted HFT order stays in the book (measured by the parameter $\gamma^H$), while keeping all the other parameters at their baseline values. 
The results of this experiment are reported in Table \ref{tab:ordercancellation_table}. We find that a reduction in the order cancellation rate (higher $\gamma^H$, see Table \ref{tab:ordercancellation_table}) decreases market volatility and the number of flash-crash episodes \footnote{We also carried out simulations for $\gamma^H>20$. The above patterns are confirmed. Interestingly, flash crashes completely disappear when the order cancellation rate is very low.}. This outcome stems from the lower aggressiveness of HF traders' strategies as order cancellation rates decrease, \textit{i.e.} the liquidity-fishing effect becomes weaker. In contrast, the duration of flash crashes is inversely related to the order cancellation rate (cf. fourth column of Table \ref{tab:ordercancellation_table}). This outcome can be explained by the order-memory effect. As $\gamma^H$ increases, the bid and ask quotes posted by HF agents stay longer in the LOB thus raising the number of contracts traded at prices close to the crash one. In turn, this hinders the recovery of the market price.

\section{Concluding Remarks}\label{Section:Conclusions}

We developed an agent-based model of a limit-order book (LOB) market to study how the interplay between low- and high-frequency traders shapes asset price dynamics and eventually leads to flash crashes. In the model, low-frequency traders can switch between fundamentalist and chartist strategies. High-frequency (HF) traders employ \textit{directional} strategies to exploit the order book information released by low-frequency (LF)
agents. In addition, LF trading rules are based on \textit{chronological} time, whereas HF ones are framed in \textit{event} time, \textit{i.e.} the activation of HF traders endogenously depends on past price fluctuations. 

We showed that the model is able to replicate the main stylized facts of financial markets. Moreover, the presence of HF traders generates periods of high market volatility and sharp price drops with statistical properties akin to the ones observed in the empirical literature. In particular, the emergence of flash crashes is explained by the interplay of three factors: i) HF traders causing periods of high illiquidity represented by large bid-ask spreads; ii) the synchronization of HF traders' orders on the sell-side of the LOB; iii) the concentration of LF traders on the buy side of the book. 

Finally, we have investigated the recovery phases that follow price-crash events, finding that HF traders' order cancellations play a key role in shaping asset price volatility and the frequency as well as the duration of flash crashes. Indeed, higher order cancellation rates imply higher market volatility and a higher occurrence of flash crashes. However, we establish that they speed up the recovery of market price after a crash. Our results suggest that order cancellation strategies of HF traders cast more complex effects than thought so far, and that regulatory policies aimed at curbing such practices (\textit{e.g.,} the imposition of cancellation fees, see also Ref.~\cite{AitSahaliaSaglam}) should take such effects into account.

Our model could be extended in at least three ways. First, we have made several departures from the zero-intelligence framework, which has been so far the standard in agent-based models of HFT. However, one can play with agents' strategic repertoires even further. For example, one could allow HF traders to switch between sets of different strategies with increasing degrees of sophistication. Second, we have considered only one asset market in the model. However, taking into account more than one market would allow one to consider other relevant aspects of HFT and flash crashes such as the possible emergence of systemic crashes triggered by sudden and huge price drops in one market \cite{securities2010findings}. In addition, another salient feature of HF traders is the ability to rapidly process and profit from the information coming from different markets \textit{i.e.,} latency arbitrage strategies \cite{wah2013latency}. Third, and finally, one could employ the model as a test-bed for a number of policy interventions directed to affect high-frequency trading and therefore mitigating the effects of flash crashes. Besides the aforementioned example of order cancellation fees, the possible policy list could include measures such as the provision of different types of trading halt facilities and the introduction of a tax on high-frequency transactions.

\clearpage \newpage \onecolumngrid

\begin{table}[h!]
\caption{Parameters values in the baseline scenario\label{tab:paramValues} }
\begin{tabular}{lll}
\hline \hline\noalign{\smallskip}
Description & Symbol & Value  \\
\noalign{\smallskip}\hline\noalign{\smallskip}
Monte Carlo replications & $MC$ & $50$ \\
Number of trading sessions & $T$ & $1,200$ \\
Number of low-frequency traders  & $N_{L}$ & $10,000$ \\
Number of high-frequency traders & $N_{H}$ & $100$ \\
LF traders' trading frequency mean& $\theta$ & $20$ \\
LF traders' min and max trading frequency & $[\theta_{min},\theta_{max}]$ & [10,40] \\
Chartists' order size parameter & $\alpha_{c}$ & 0.04 \\
Chartists' shock standard deviation &  $\sigma^{c}$ & 0.05 \\
Fundamentalists' order size parameter & $\alpha_{f}$ & 0.04 \\
Fundamentalists' shock standard deviation &  $\sigma^{f}$ & 0.01 \\
Fundamental value shock standard deviation & $\sigma^{y}$ & 0.01 \\
Price drift parameter & $\delta$ & 0.0001 \\
LF traders' price tick standard deviation & $\sigma^{z}$ & 0.01 \\
LF traders' intensity of switching  & $\zeta$ & 1 \\
LF traders' resting order periods  & $\gamma^{L}$ & 20 \\
HF traders' resting order periods  & $\gamma^{H}$ & 1 \\
HF traders' activation threshold distribution support & $[\eta_{min},\eta_{max}]$ & [0,0.2] \\
Market volumes weight in HF traders' order & $\lambda$ & $0.625$ \\
size distribution &&\\
HF traders' order price distribution support & $[\kappa_{min},\kappa_{max}]$ & [0,0.01] \\
\noalign{\smallskip}\hline \hline
\end{tabular}
\end{table}

\twocolumngrid

\end{document}